\documentclass[a4paper,11pt]{article}
\usepackage{pos}
\usepackage{hyperref}
\usepackage{comment}
\usepackage{booktabs}
\newcommand{\varA}[1]{{\operatorname{#1}}}
\usepackage{adjustbox}

\title{Estimating QCD uncertainties on antiproton spectra from dark-matter annihilation}
\ShortTitle{Estimating QCD uncertainties on antiproton spectra from dark-matter annihilation}

\author*[a]{Adil Jueid}
\author[b]{Jochem Kip}
\author[c]{Roberto Ruiz de Austri}
\author[d]{Peter Skands}

\affiliation[a]{Quantum Universe Center, Korea Institute for Advanced Study, 02455 Seoul, Republic of Korea}
\affiliation[b]{Institute for Mathematics, Astrophysics and Particle Physics, Radboud University
Nijmegen, Heyendaalseweg 135, Nijmegen, the Netherlands}
\affiliation[c]{Instituto de F\'isica Corpuscular, IFIC-UV/CSIC, Valencia, Spain}
\affiliation[d]{School of Physics and Astronomy, Monash University, VIC-3800, Australia}

\emailAdd{adiljueid@kias.re.kr}
\emailAdd{jochem.kip@ru.nl}
\emailAdd{rruiz@ific.uv.es}
\emailAdd{peter.skands@monash.edu}

\abstract{In this talk, we discuss the physics modeling of antiproton spectra arising from dark matter (DM) annihilation or decay in a model-independent manner. The modeling of antiproton spectra contains some intrinsic uncertainties related to QCD parton showers and hadronisation of baryons. We briefly assess the sources of these uncertainties and their impact on antiproton energy spectra for a few selected DM scenarios. The results are provided in tabulated form for future analyses.}

\FullConference{%
  Computational Tools for High Energy Physics and Cosmology - CompTools2021 \\
  22 -- 26 November, 2021\\
  Institut de Physique des 2 Infinis (IP2I), Lyon, France}

\begin{document}
\maketitle

\section{Introduction}
\label{sec:introduction}

Weakly interacting massive particles (WIMPs) embedded in various new physics models beyond the Standard Model (SM) are expected to annihilate into SM particles which, after a complex sequence of processes, lead to stable final-state particles such as photons, antineutrinos, positrons or antiprotons. These stable final-state objects may leave footprints in experiments such as the Fermi Large Area Telescope (Fermi-LAT), IceCube or the Alpha Magnetic Spectrometer (AMS). Soon after the discovery of the secondary cosmic ray antiprotons \cite{Golden:1979bw, Buffington:1981zz}, a mild excess over the backgrounds was reported. This excess has been explained shortly afterwards by considering the DM candidate to be a massive photino in a supersymmetric model \cite{Silk:1984zy, Stecker:1985jc}. The excess seems to be still there despite the unprecedented precision on the measurements of the antiproton-to-proton ratio by the AMS--02 collaboration \cite{AMS:2016oqu}. The statistical uncertainties on the measurement itself is subleading now thanks to the large amount of data collected for a four-year period between 2011 and 2015 and over the rigidity range of $1$--$450$~GV. Based on this measurement, several groups have reported an excess over the SM backgrounds has been observed in the rigidity range of $10$--$20$ GV \cite{Cuoco:2016eej,Cui:2016ppb,Cuoco:2017rxb,Reinert:2017aga,Cui:2018klo,Cuoco:2019kuu,Cholis:2019ejx,Lin:2019ljc,Abdughani:2021pdc,Hernandez-Arellano:2021bpt,Biekotter:2021ovi}. In this case, a dark matter with mass around $\sim 60$--$200$~GeV and thermal annihilation cross section of about $\sim 10^{-26}~{\rm cm}^2~s^{-1}$ seems to explain the AMS--02 excess. Interestingly, models with dark matter having almost similar properties are able to address the so-called gamma-ray Galactic Center Excess (GCE) \cite{Goodenough:2009gk, Vitale:2009hr, Hooper:2010mq, Gordon:2013vta, Hooper:2011ti, Daylan:2014rsa, Zhou:2014lva, Calore:2014xka, Abazajian:2014fta,Zhou:2014lva, Caron:2015wda, vanBeekveld:2016hbo, Butter:2016tjc, Karwin:2016tsw, Achterberg:2017emt}. An important feature of these analyses is that modeling of uncertainties may play a very important role not only in the discovery reach of dark matter in indirect detection experiments but also in post-discovery studies. However, a proper treatment of all the theoretical and systematic uncertainties on the antiproton flux is usually overlooked in the literature. It was found that proper treatment of the systematic uncertainties and their correlations can have drastic consequences on the AMS--02 excess \cite{diMauro:2014zea,Kappl:2014hha,Kachelriess:2015wpa,Winkler:2017xor,Korsmeier:2018gcy, Heisig:2020nse}. In the process of dark matter annihilation, antiprotons are solely produced from the QCD jet fragmentation. The process of jet fragmentation which occurs at the scale of the proton mass can not be solved from first principles but using phenomenological models or parametric fits. Hadronisation is usually performed using multi-purpose Monte Carlo event generators. We note that there are two such models: the string model \cite{Artru:1974hr,Andersson:1983ia} (used in \textsc{Pythia}~8 \cite{Sjostrand:2014zea}) and the cluster model (used in \textsc{Herwig}~7 \cite{Bellm:2015jjp} and \textsc{Sherpa}~2 \cite{Gleisberg:2008ta}). The hadronisation models depend, in principle, on many parameters that can be constrained from data at colliders such as e.g. LEP. The modeling of the antiproton flux, therefore, contains some uncertainties that are overlooked in the literature\footnote{Some comparisons between different Monte Carlo (MC) event generators have been carried out in \cite{Cirelli:2010xx,Cembranos:2013cfa}. We however argue that the uncertainties obtained from the envelopes of different Monte Carlo event generators are not conservative as they do not span the interval allowed by the experimental data. It was found that the differences in particle spectra predicted in different MC event generators can be observed in the tails of the spectra while there is a high level of agreement between them in the bulk of the spectra. This finding has been confirmed in a previous study where we have used the most recent and widely used MC event generators \cite{Amoroso:2018qga}.}. In a previous study, we have shown that QCD uncertainties on gamma-ray dark matter searches can be important and we provided for the first time a conservative estimate of QCD uncertainties within \textsc{Pythia}~8 \cite{Amoroso:2018qga} (a short summary can be found in \cite{Amoroso:2020mjm, Jueid:2021dlz}). The aim of this talk to discuss the QCD uncertainties on antiproton spectra following the same methods used in the previous study\footnote{This talk is based on reference \cite{Jueid:2022qjg, Jueid:2022xaz}}. To assess the QCD uncertainties on the antiproton fluxes, we first revisit the constraints from LEP measurements on the parameters of the Lund fragmentation function and discuss the differences between the various measurements of baryon spectra at the $Z$--pole. We then perform several tunings based on the baseline \textsc{Monash} tune \cite{Skands:2014pea} of the \textsc{Pythia}~8.244 event generator \cite{Sjostrand:2014zea}. We estimate the QCD uncertainties as the result of various eigentunes from the optimisation tool \textsc{Professor} 2.3.3 \cite{Buckley:2009bj} based on the measurements implemented in \textsc{Rivet} 3.1.3 \cite{Bierlich:2019rhm}. The rest of the paper is organised as follows. In section \ref{sec:physics}, we discuss the antiproton production from generic dark matter annihilation, and the relevant measurements that can be used to constrain the spectra. In section \ref{sec:tunes}, we discuss the global analysis of the fragmentation function including baryon production. We discuss the different sources of QCD uncertainties on the antiproton spectra and assess their impact for a few selected dark matter masses and annihilation channels and show for comparison the spectra of positrons in section \ref{sec:uncertainties}. We summarize our conclusions in section \ref{sec:conclusions}.

\section{Antiprotons from dark matter annihilation: Signals and constraints}
\label{sec:physics}

\subsection{Theoretical modeling of proton production}
We first study the production of antiprotons from a generic dark matter annihilation or decay process\footnote{A more detailed discussion can be found in refs. \cite{Amoroso:2018qga, Jueid:2022xaz}.}. This discussion applies in general for any dark matter decaying to hadronic final states and whose masses are above a few GeV. The process under consideration can be written schematically as follows:
\begin{eqnarray}
\chi \chi \to \underbrace{X_1 X_2 \cdots X_N}_{\varA{parton-level~particles}} \to \overbrace{\bigg(\displaystyle\prod_{i=1}^{a_1} Y_{1i} \bigg) \bigg(\displaystyle\prod_{j=1}^{a_2} Y_{2j} \bigg) \ldots \bigg(\displaystyle\prod_{z=1}^{a_N} Y_{Nz} \bigg)}^{\varA{hadron-level~final-state~objects}},
\label{eq:process}
\end{eqnarray}
where the narrow-width approximation is used to factorise the whole process into a production part $\chi \chi \to \prod_{i=1}^{N} X_i$ and a decay part $X_i \to \prod_{k=1}^{a_i} Y_{ik}$. It is understood that the decay part must contain in addition to resonant decay of heavy SM particles the process of QCD bremsstrahlung and hadronisation to have antiprotons in the final state. Coloured particles produced either directly in dark matter annihilation or from the decay of heavy resonances undergo QCD bremsstrahlung processes where additional quarks and gluons are produced. The rate of QCD bremsstrahlung processes is controlled by the value of the strong coupling constant $\alpha_S(M_Z)$. Note that the value of $\alpha_S(M_Z)$ in \textsc{Pythia}~8 is larger than $\alpha_S(M_Z)^{\overline{\rm MS}}$ by $20\%$ \cite{Skands:2010ak,Skands:2014pea}.  Finally, at a scale $Q_{\rm IR} \simeq \mathcal{O}(1)~{\rm GeV}$, any coloured particle must hadronise to produce a set of colourless hadrons. This process, called fragmentation is modeled within \textsc{Pythia}~8 with the Lund string model \cite{Andersson:1983ia, Sjostrand:1982fn, Sjostrand:1984ic}. The description of the hadronisation process is achieved in the \emph{fragmentation function}, $f(z)$, which gives the probability for a hadron to take a fraction $z \in [0, 1]$ of the remaining energy at each step of the (iterative) string fragmentation process. The general form can be written as 
\begin{equation}
        f(z,m_{\perp h}) \propto N \frac{(1-z)^a}{z}\exp\left(\frac{-b m_{\perp h}^2}{z}\right)~,
\label{eq:fz}
\end{equation}
where $N$ is a normalisation constant that guarantees the distribution to be normalised to unit integral, and $m_{\perp h}= \sqrt{m_h^2 + p_{\perp h}^2}$ is called the ``transverse mass'', with $m_h$ the mass of the produced hadron and $p_{\perp h}$ its momentum transverse to the string direction, $a$ and $b$ are tunable parameters. It was found in ref. \cite{Amoroso:2018qga} that the $a$ and $b$ parameters are highly correlated. Therefore, a new parametrisation of the fragmentation function exists for which the $b$ parameter is replaced by $\langle z_\rho \rangle$ which represents the average longitudinal momentum fraction taken by mainly the $\rho$ mesons and which is computed at the initialisation state. Baryon production in \textsc{Pythia8} is controlled by an additional parameter $a_{\rm Diquark}$ that represents the rate for diquark production in the fragmentating strings. In this case, the $a$ parameter in $f(z)$ is modified as $a \to a + a_{\rm Diquark}$. table \ref{tab:ranges} shows the parameters of the fragmentation function in \textsc{Pythia}~8, their default values in the baseline \textsc{Monash} tune and their allowed intervals.

\begin{table}[t!]
\setlength\tabcolsep{8pt}
  \begin{center}
    \begin{tabular}{llcll}
      \toprule
      parameter & \textsc{Pythia8} setting & Variation range & \textsc{Monash} & Tune-3D \cite{Amoroso:2018qga} \\
      \midrule
      $\sigma_{\perp}$~[GeV] & \verb|StringPT:Sigma|   & 0.0 -- 1.0 & 0.335 & 0.3174 \\
      $a$              & \verb|StringZ:aLund|    & 0.0 -- 2.0 & 0.68 & 0.5999 \\
      $b$              & \verb|StringZ:bLund|    & 0.2 -- 2.0 & 0.98 & -- \\
      $\left<z_\rho\right>$       & \verb|StringZ:avgZLund| & 0.3 -- 0.7 & (0.55) & 0.5278 \\
      $a_{\rm DiQuark}$ & \verb|StringZ:aExtraDiquark| & 0.0 -- 2.0 & 0.97 & 0.97 \\
      \bottomrule
    \end{tabular}
  \end{center}
  \caption{\label{tab:ranges} Parameter ranges used for the \textsc{Pythia} 8 tuning,
    and their corresponding values in the \textsc{Monash} tune \cite{Skands:2014pea} and in a tune performed in \cite{Amoroso:2018qga}.}
\end{table}

\begin{figure}[!t]
    \centering
    \includegraphics[width=0.495\linewidth]{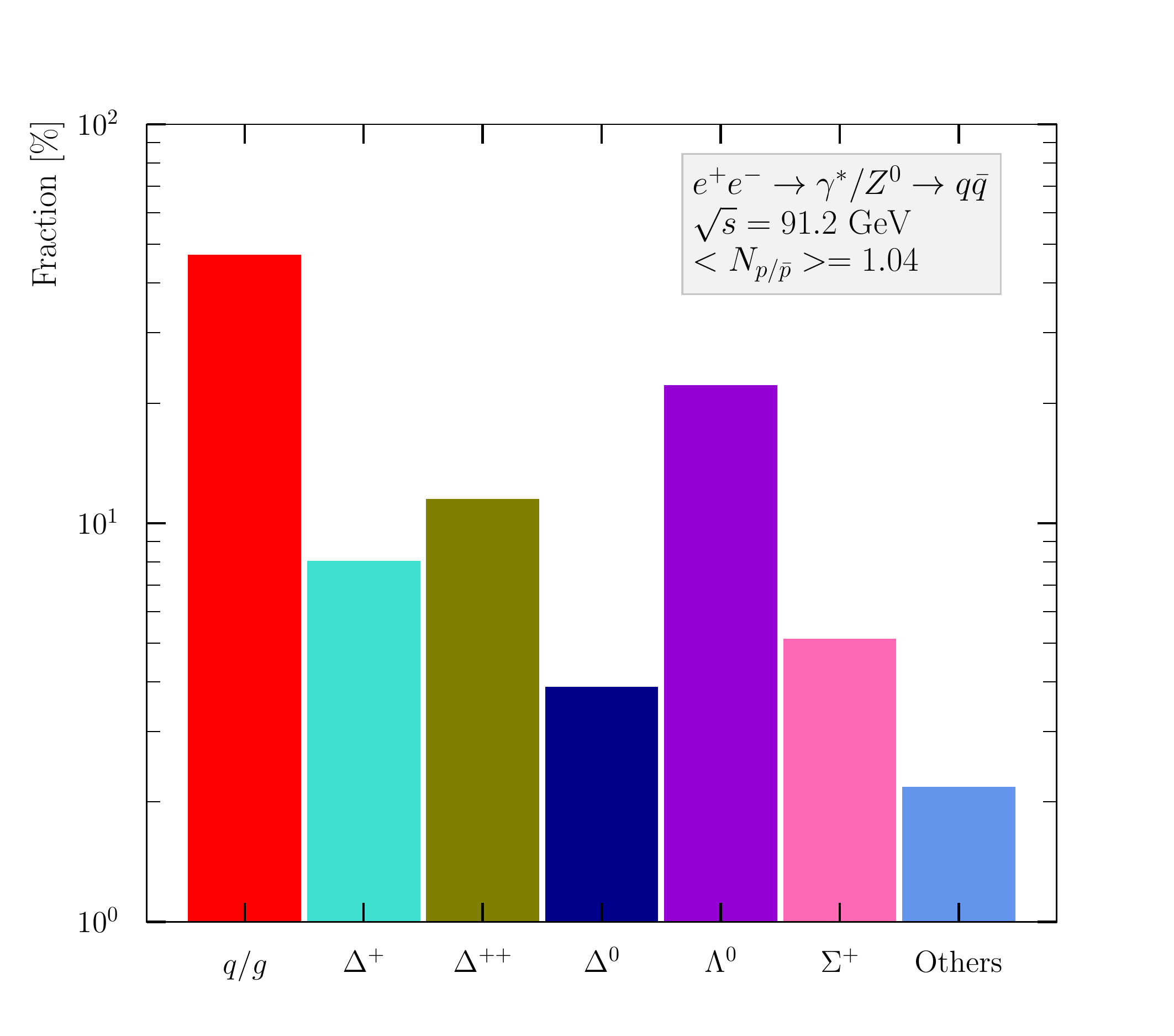}
    \hfill
    \includegraphics[width=0.495\linewidth]{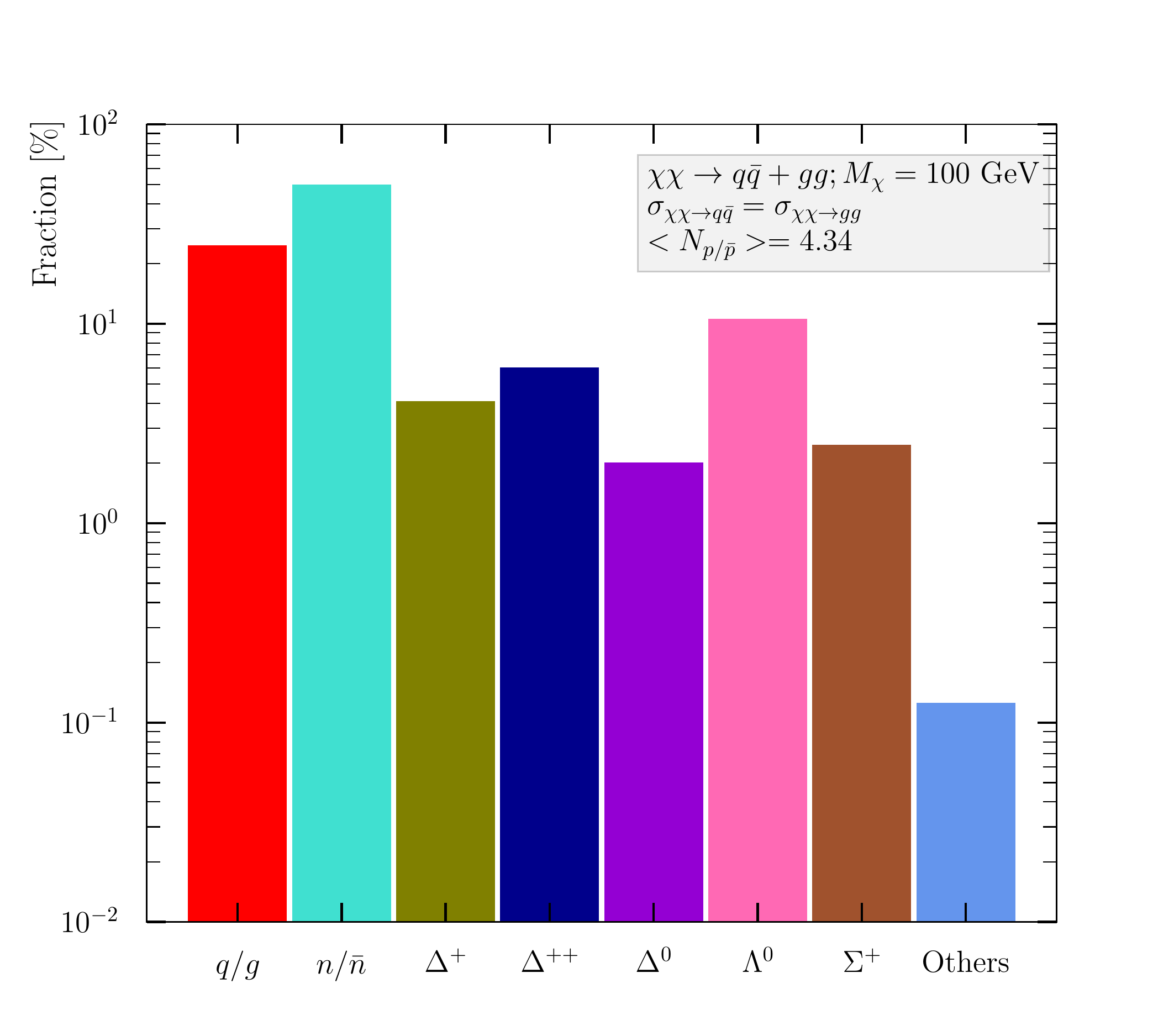}
    \vspace{-0.5cm}
    \caption{The mean contribution to $p/\bar{p}$ production in $e^+ e^- \to \gamma^*/Z^0 \to q\bar{q}$ at $\sqrt{s} = 91.2~{\rm GeV}$ ({\it left}) and in dark matter annihilation into $q\bar{q}$ for $M_{\chi} = 100~{\rm GeV}$~({\it right}). Here, one shows $p/\bar{p}$ produced from QCD fragmentation, neutron decay, $\Delta^+$, $\Delta^{++}$, $\Delta^0$, $\Lambda^0$, $\Sigma^+$ and from other baryons.}
    \label{fig:protonsOrigin}
\end{figure}

In figure \ref{fig:protonsOrigin}, we show the origin of protons in both $e^+ e^- \to q\bar{q}$ at LEP ({\it left}) and in dark matter annihilation into $q\bar{q}$ for $M_{\chi} = 100~{\rm GeV}$~({\it right}). In general, (anti-)protons can be split into two categories: \emph{(i)} primary (anti-)protons produced directly from the string fragmentation of quarks and gluons and \emph{(ii)} secondary (anti-)protons produced from the decay of heavier baryons. 

\subsection{Experimental constraints}

From the discussion in the previous subsection, it is clear that the modeling of antiprotons will be improved if one includes all the relevant measurements of proton spectra performed at LEP. Besides the measurements of the proton spectrum itself, one may expect some improvements from measurements of the spectra of the $\Lambda^0$ baryons as well. The $\Lambda^0$ baryons are dominant sources of secondary protons at LEP~(about $22\%$). There is a strong correlation between the spectra of (anti-)protons and of $\Lambda^0$ baryons since all the $\Lambda^0$ baryons are reconstructed from their decays into $\pi p$. We do not expect significant improvements from measurements of other baryons such as $\Delta^{++}$ or $\Sigma^{\pm}$ as due to the limited precision on their differential rates. Therefore, the main constraining observables in this study will consist of a set of measurements of $\Lambda$ and $p/\bar{p}$ energy--momentum distributions. To guarantee a good agreement with the results of the previous study \cite{Amoroso:2018qga} (including an acceptable modelling of overall event properties), we also include measurements of meson spectra, event shapes and particle multiplicities. We have considered eight constraining measurements of proton and $\Lambda^0$ reported on by \textsc{Aleph} \cite{Barate:1996fi,Barate:1999gb}, \textsc{Delphi} \cite{Abreu:1993mm, Abreu:1995cu, Abreu:1998vq} and \textsc{Opal} \cite{Akers:1994ez,Alexander:1996qj}. 

\begin{figure}[!t]
    \centering
    \includegraphics[width=0.49\linewidth]{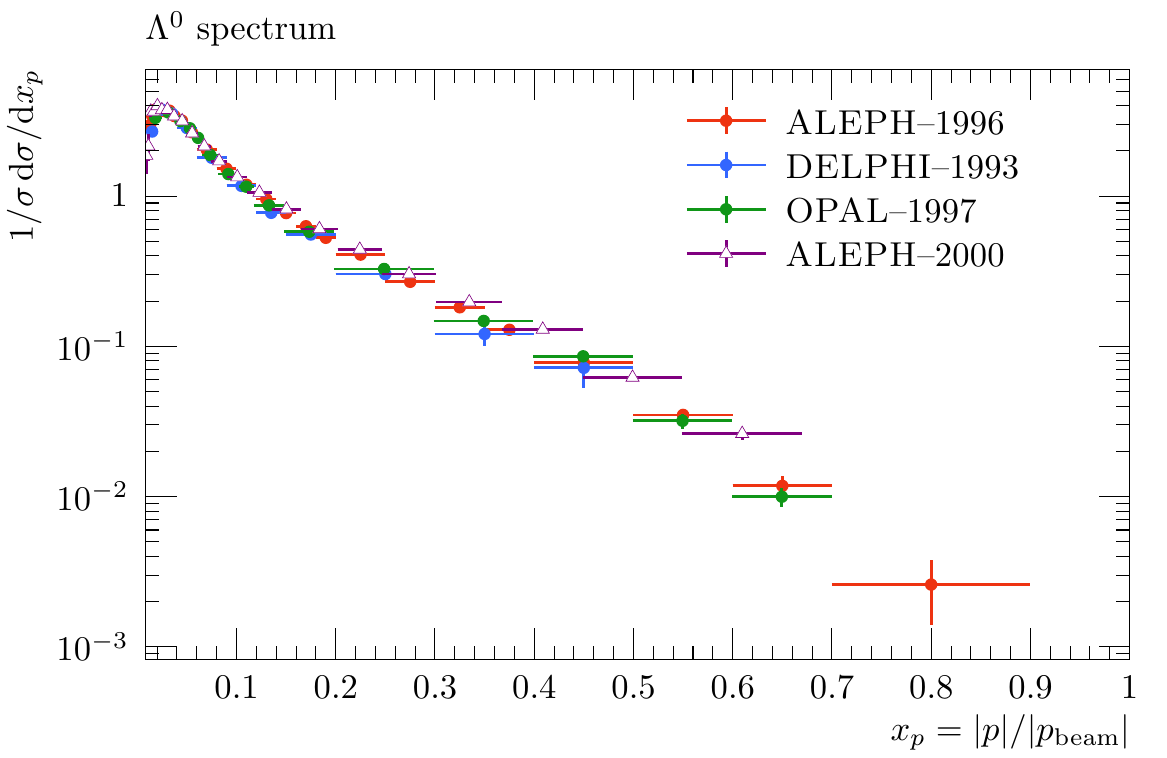} \hfill
    \includegraphics[width=0.49\linewidth]{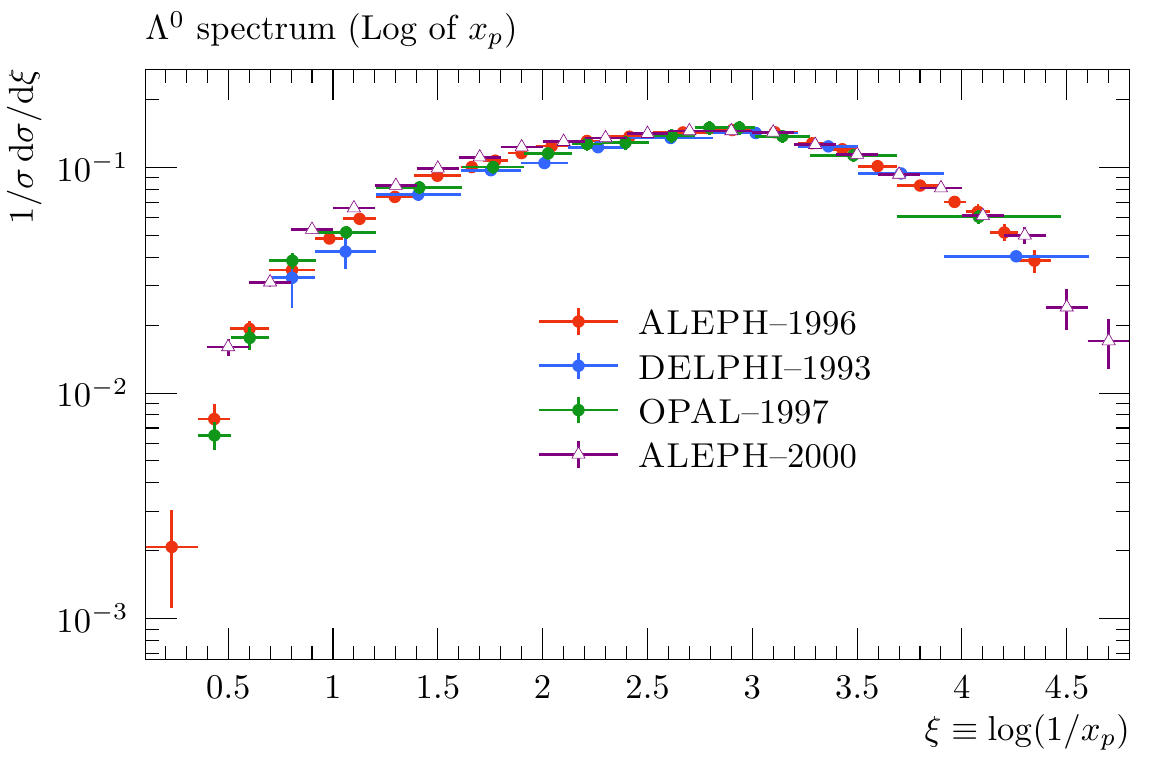}
    \vfill
    \includegraphics[width=0.49\linewidth]{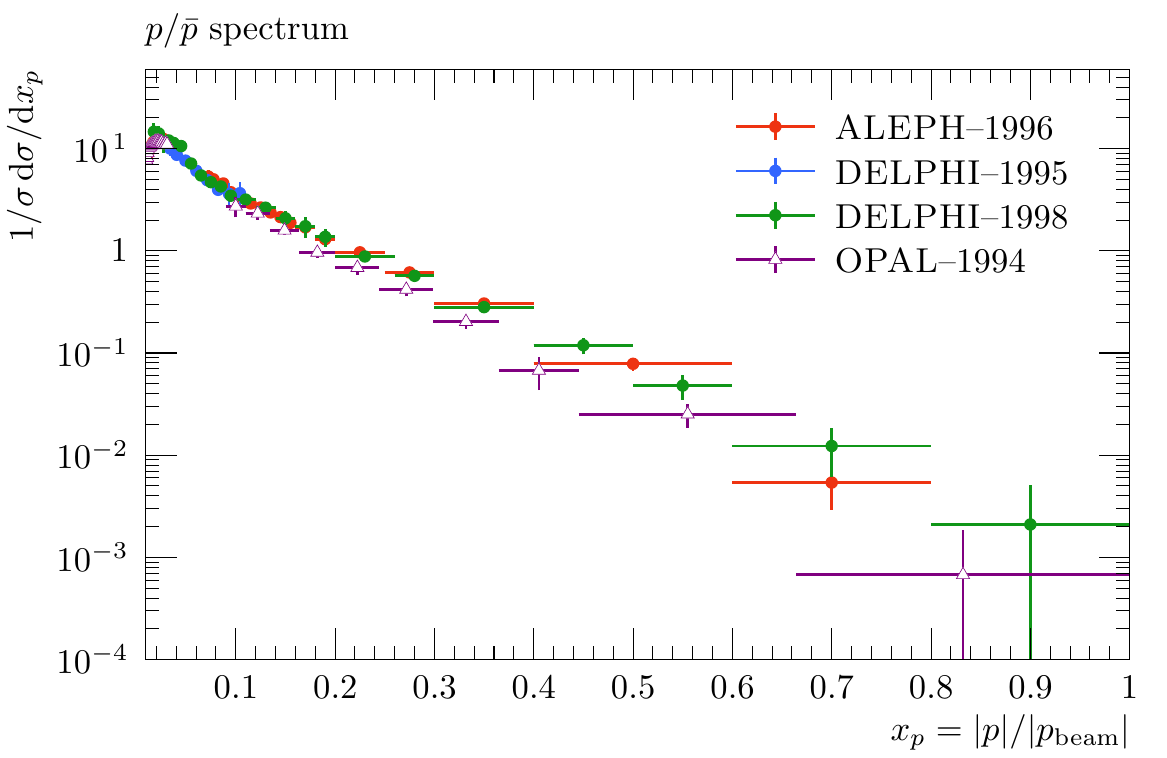}
    \hfill
    \includegraphics[width=0.49\linewidth]{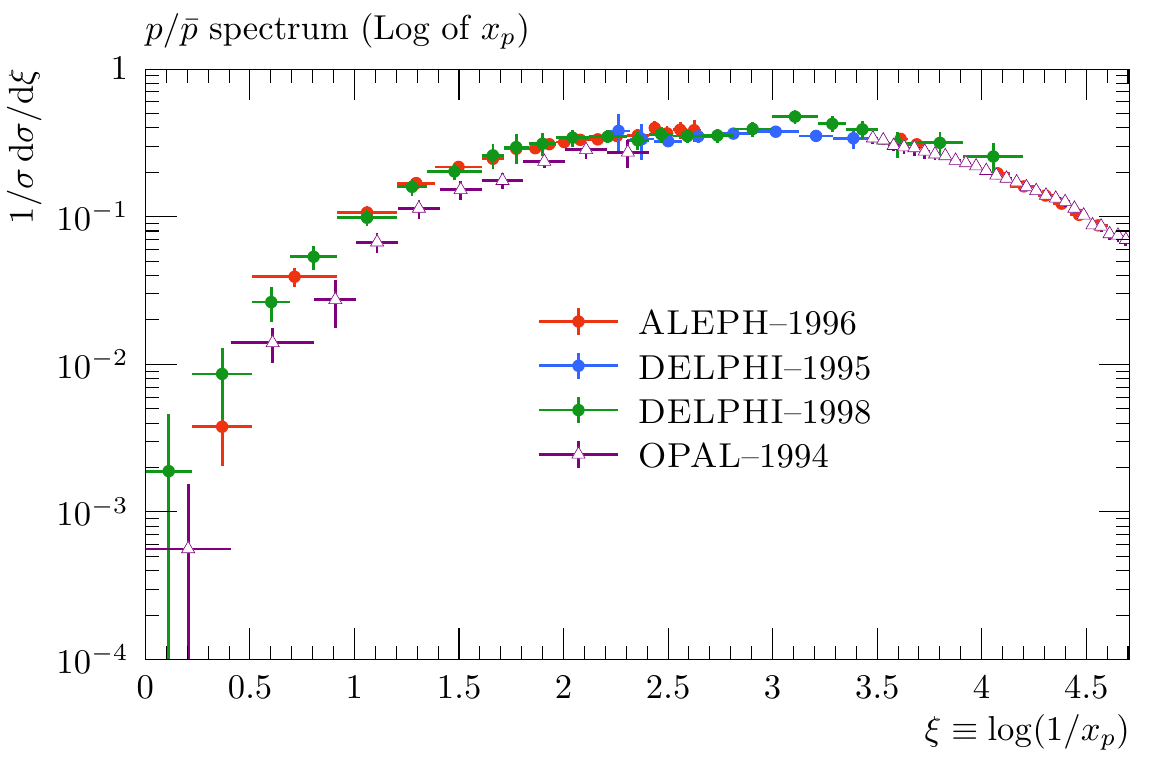}
    \caption{Comparison between the different measurements of the $\Lambda^0$ scaled momentum ({\it upper panels}) and of the proton scaled momentum ({\it lower panels}). Data is taken from \cite{Abreu:1993mm, Barate:1996fi, Akers:1994ez, Abreu:1995cu, Abreu:1998vq, Abe:1998zs, Barate:1999gb, Abe:2003iy}.}
    \label{fig:comparisons}
\end{figure}  

In figure~\ref{fig:comparisons}, we show the comparisons between the different experimental measurements of $\Lambda^0$~spectrum and $p/\bar{p}$ spectrum. We can see that there are some inconsistencies between the measurement of the scaled momentum of $p/\bar{p}$ performed by \textsc{Opal} and the other experiments for $x_p > 0.1$ (the \textsc{Opal} result is below all the others). Furthermore, the old \textsc{Delphi} measurement (blue) of $p/\bar{p}$ momentum is inconsistent with the new one (green) for few bins of $\xi \simeq 3$--$3.2$. Note that both these \textsc{Delphi}~measurements cover the hole left by \textsc{Aleph}--1996. Finally, the \textsc{Delphi}--1993 measurement of $\Lambda^0$ scaled momentum seems to be inconsistent with the others for for $\xi < 1.1$ (the discrepancy is mild as compared to the proton case). We do not include \textsc{Opal}--1994 measurement as it is inconsistent with all the other measurements for $x_p > 0.1$ and \textsc{Delphi}--1995 as it was superseded by \textsc{Delphi}--1998 one. 

\section{Tunes of the fragmentation function}
\label{sec:tunes}

\subsection{Phenomenological setup}

This study aims to perform further retunings of the fragmentation function using \textsc{Pythia8} version 8.244 \cite{Sjostrand:2014zea} assuming the \textsc{Monash} tune \cite{Skands:2014pea} as our baseline. The different measurements used in our tunings are implemented in the validation package \textsc{Rivet} version 3.1.3 \cite{Bierlich:2019rhm}. Both Frequentist and Bayesian tunings are performed using the optimisation tool \textsc{Professor} version 2.3.3 \cite{Buckley:2009bj} and \textsc{MultiNest} \cite{Feroz:2008xx}. Analytical expressions for the physical dependence of the observables on the different parameters are derived by fitting the Monte Carlo predictions to a set of points in the parameter space. In this study we assume a fourth-order polynomial interpolation to represent the true response of \textsc{Pythia}~8, {\it i.e.} 

\begin{eqnarray}
    f_{(b)}(\{p_i\}) = \alpha_0^{(b)} + \sum_{i=1}^4 \beta_i^{(b)} p_i + \sum_{i,j=1}^4 \gamma_{ij}^{(b)} p_i p_j + \sum_{i,j,k = 1}^4 \delta_{ijk}^{(b)} p_i p_j p_k + \sum_{i,j,k,\ell=1}^4 \epsilon_{ijk\ell}^{(b)} p_i p_j p_k p_\ell,
    \label{eq:interp}
\end{eqnarray}
with $\alpha, \beta, \gamma, \delta,~{\rm and}~ \epsilon$ are the polynomial coefficients determined in the fit and $\{p_i\} = \{a, \langle z_\rho \rangle, \sigma_\perp, a_{\rm Diquark} \}$ are the parameters of the Lund fragmentation function (defined in table \ref{tab:ranges}). The best-fit points for the parameters are determined by a standard $\chi^2$-minimisation method (\textsc{Minuit} \cite{James:1975dr}) implemented in \textsc{Professor} and which uses the analytical polynomial interpolations defined in equation \ref{eq:interp}.

The goodness-of-fit per number of degrees of freedom is defined as 
\begin{equation}
 \chi^2 = \frac{1}{\sum_{\mathcal{O}} \omega_\mathcal{O} |b \in \mathcal{O}| - N_{\rm parameters}} \sum_{\mathcal{O}}
 \omega_\mathcal{O} \sum_{b\in \mathcal{O}} \bigg(\frac{f_{(b)}(\{p_i\}) - \mathcal{R}_b}{\Delta_b}\bigg)^2,
\label{eq:GoF}
\end{equation}
where $\mathcal{R}_b$ is the central value for the experimental measurement $\mathcal{O}$ at a bin $b$, $f_{(b)}(\{p_i\})$ is the analytical expression of the response function which is a polynomial of the parameters, and $\Delta_b$ is the total error which is quadratic sum of the statistical MC errors, the experimental errors and a flat $5\%$ theory errors:
\begin{eqnarray}
\Delta_b = \sqrt{\sigma_{b, \rm exp}^2 + \sigma_{b, {\rm MC}}^2 + \sigma_{b, {\rm th}}^2},
\end{eqnarray}
with $\sigma_{b, {\rm th}} = 0.05 \times f_{(b)}(\{p_i\})$.

\subsection{Results}

\begin{table}[t!]
\setlength\tabcolsep{3pt}
  \begin{center}
    \begin{tabular}{lcccccc}
      \toprule
      Tune     & \verb|aLund| & \verb|avgZLund| & \verb|sigma| & \verb|aExtraDiquark| & \verb|bLund| &  $\chi^2/N_{\rm df}$\\
      \midrule
      \textsc{Aleph}  & $0.758\substack{+0.074\\ -0.074}$ & $0.541\substack{+0.007\\ -0.007}$ & $0.297\substack{+0.005\\-0.005}$ & $1.218\substack{+0.358\\-0.358}$ & $1.040$ & $116.22/296$ \\
      \textsc{Delphi} & $0.358_{-0.054}^{+0.054}$ & $0.497_{-0.007}^{+0.007}$ & $0.287_{-0.006}^{+0.006}$ & $0.782_{-0.298}^{+0.298}$ & $0.533$
      &  $144.37/268$ \\
      \textsc{L3}     &  $0.478_{-0.063}^{+0.063}$  & $0.557_{-0.006}^{+0.006}$ &  $0.315_{-0.007}^{+0.007}$ & $1.998_{-0.049}^{+0.049}$ & $0.897$ & $84.70/140$ \\
      \textsc{Opal}   & $0.588_{-0.086}^{+0.086}$ & $0.536_{-0.005}^{+0.005}$ &     $0.300_{-0.005}^{+0.005}$    &  $1.998_{-0.204}^{+0.204}$ & $0.872$ &     $53.54/136$ \\
      \midrule
      \textsc{Combined} &       $0.601_{-0.038}^{+0.038}$ & $0.540_{-0.004}^{+0.004}$ &  $0.307_{-0.002}^{+0.002}$ & $1.671_{-0.196}^{+0.196}$ & $0.897$ &     $676.69/852$ \\
      \bottomrule
    \end{tabular}
  \end{center}
    \caption{\label{tab:experiments} Results of the tunes performed 
    separately to all the considered measurements from a given experiment.} 
\end{table}

\begin{figure}[!tbp]
    \centering
    \includegraphics[width=0.75\textwidth]{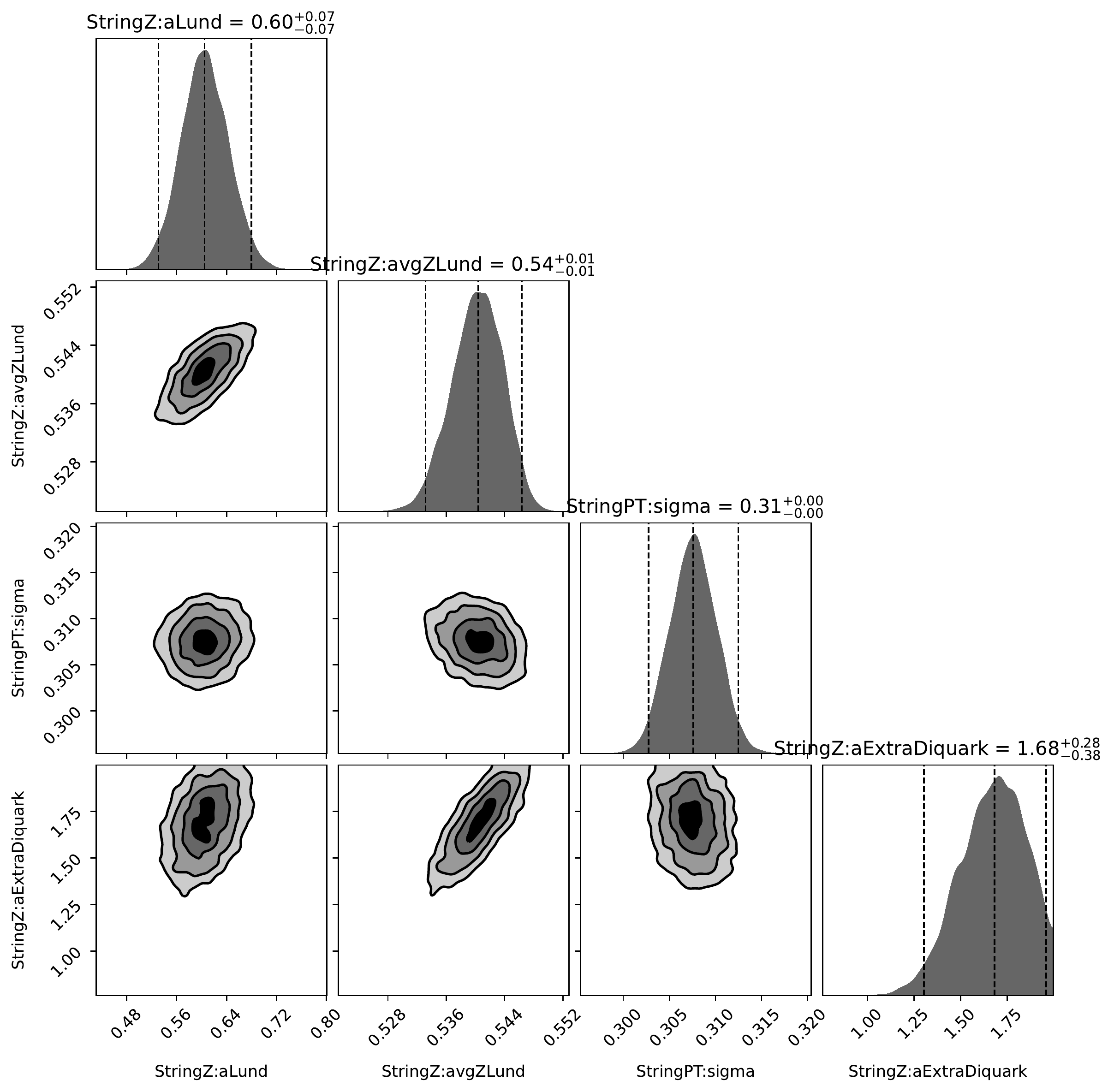}
    \caption{One- and two-dimensional marginalized posterior distributions for the uni-modal four-dimensional parameter space fit. Here, the contours show the $68\%$ and $95\%$ Bayesian credible intervals.}
    \label{fig:bayesian:results:combined}
\end{figure}

The results of the fragmentation function tunes are shown in table \ref{tab:experiments} for individual experiments and their combinations. We can see that the best-fit points of \texttt{StringZ:aLund}, \texttt{StringZ:avgZLund} and \texttt{StringPT:sigma} are consistent with the results of a previous study \cite{Amoroso:2018qga}. On the other hand, the best-fit point of \texttt{StringZ:aExtraDiquark} is larger than what we found in the one-dimensional parameter space tune. Note that this value is driven by the results from two experiments: \textsc{L3} and \textsc{Opal}. In figure \ref{fig:bayesian:results:combined} we show one- and two-dimensional marginalized posterior distributions for the unimodular four-dimensional parameter space fit along with the $68\%$ and $95\%$ Bayesian credible intervals. We have also checked that these results do not correspond to flat directions as the best-fit point of \texttt{StringZ:aExtraDiquark} does not depend on the maximum value allowed in the scan. We finally note that the these tune results give fairly good agreement with data and the results are competitive with the baseline \textsc{Monash} tune.

\section{QCD uncertainties on antiproton spectra}
\label{sec:uncertainties}

\subsection{Perturbative uncertainties}
The perturbative uncertainties are split into catergories: scale variation uncertainties and non-singular (hard terms) uncertainties. The first class of the shower uncertainties are estimated by varying the renormalisation scale by a factor of two in each direction with respect to the nominal scale choice. To guarantee that scale variations are as  conservative as possible, a number of modifications on the naive scale variations are done by adding some NLO compensation terms to absorb large corrections \cite{Mrenna:2016sih}. The variations of the non-universal hard components of the DGLAP kernels are also possible with the new automated formalism. We close this discussion by noting that Matrix-Element Corrections (MECs), switched on by default in \textsc{Pythia}~8, lead to very small variations of the non-singular terms of the DGLAP splittings for $Z^0$ decays. It was found that switching off these corrections would lead to comprehensively larger envelopes \cite{Mrenna:2016sih}.

\subsection{Fragmentation function uncertainties}

\begin{table}[!h]
 \setlength\tabcolsep{4pt}
 \begin{center}
\begin{adjustbox}{max width=\textwidth}
  \centering
    \begin{tabular}{l c c c c}
    \toprule
    \toprule
Tune	& \texttt{StringZ:aLund} &	\texttt{StringZ:avgZLund} & \texttt{StringPT:sigma}	& \texttt{StringZ:aExtraDiquark} \\
\toprule
Central	& $0.601$	& $0.540$ &	$0.307$	& $1.671$ \\
\toprule
\multicolumn{5}{l}{\textit{$2\sigma$ eigentunes}} \\
\toprule
Variation $1^+$	& $0.609$ & $0.542$ &  $0.307$ &    $1.775$ \\
Variation $1^-$	& $0.591$ & $0.538$ &  $0.307$ &	$1.558$ \\
Variation $2^+$	& $0.501$ &	$0.535$ &  $0.306$ &	$1.679$ \\
Variation $2^-$	& $0.700$ & $0.544$ &  $0.308$ &    $1.662$ \\
Variation $3^+$	& $0.597$ &	$0.609$ &  $0.333$ &	$1.670$ \\
Variation $3^-$	& $0.603$ &	$0.474$ &  $0.283$ &	$1.671$ \\
Variation $4^+$	& $0.601$ & $0.478$ &  $0.475$ &	$1.672$ \\
Variation $4^-$	& $0.600$ & $0.581$	&  $0.197$ &    $1.669$ \\
\toprule
\bottomrule
\end{tabular}
\hspace{0.2cm}
\end{adjustbox}
\end{center}
    \caption{The Hessian variations (eigentunes) for the nominal tune including all the measurements performed by \textsc{Aleph}. The variations correspond to $\Delta \chi^2 = 4$~($95\%$ CL) with $\Delta \chi^2$ is defined as $\Delta \chi^2 \equiv \chi^2_{\rm var} - \chi^2_{\rm min}$.}
    \label{tab:eigentunes}
\end{table}

The \textsc{Professor} toolkit provides an estimate of the uncertainties on the fitted parameters through the Hessian method (also known as eigentunes). This method consists of a diagonalisation of the $\chi^2$ covariance matrix near the minimum which results in building a set of $2 \cdot N_{\rm params}$ variations. These variations are then obtained as corresponding to a fixed change in the goodness-of-fit measure which is found by imposing a constraint on the maximum variation, defined as a hypersphere with maximum radius of $T$ (defined as the tolerance), i.e. $\Delta \chi^2 \leq T$. Therefore one can define the $\Delta \chi^2$ to match a corresponding confidence level interval; {\it i.e.} one-sigma variations are obtained by requiring that $\Delta \chi^2 \simeq N_{\rm df}$ where $N_{\rm df}$ is the number of degrees-of-freedom. This approach allows for a conservative estimate of the uncertainty if the event generator being used has a good agreement with data (which is usually quantified by $\chi^2_{\rm min}/N_{\rm df} \leq 1$) and the resulting uncertainties provide a good coverage of the errors in the experimental data. To enable for this situation, we have added an extra $5\%$ uncertainty to the MC predictions for all the observables and bins which already implied a $\chi^2/N_{\rm df} \leq 1$ in our fits as depicted in table \ref{tab:experiments}. On the other hand, we enable for large uncertainties by considering not only the one-sigma eigentunes but also the two-sigma eigentunes (correspond to $\Delta \chi^2/N_{\rm df} = 4$) and the three-sigma eigentunes (correspond to $\Delta \chi^2/N_{\rm df} = 9$). The three-sigma eigentunes provide a very good coverage of all the experimental uncertianties in the data for meson and baryon spectra but results in unreasonably large uncertainties that overshot the experimental errors for {\it e.g.} event shapes or jet rates.  The variations corresponding to the two-sigma eigentunes are shown in table \ref{tab:eigentunes}. \\

\begin{figure}[!t]
    \centering
    \includegraphics[width=0.49\linewidth]{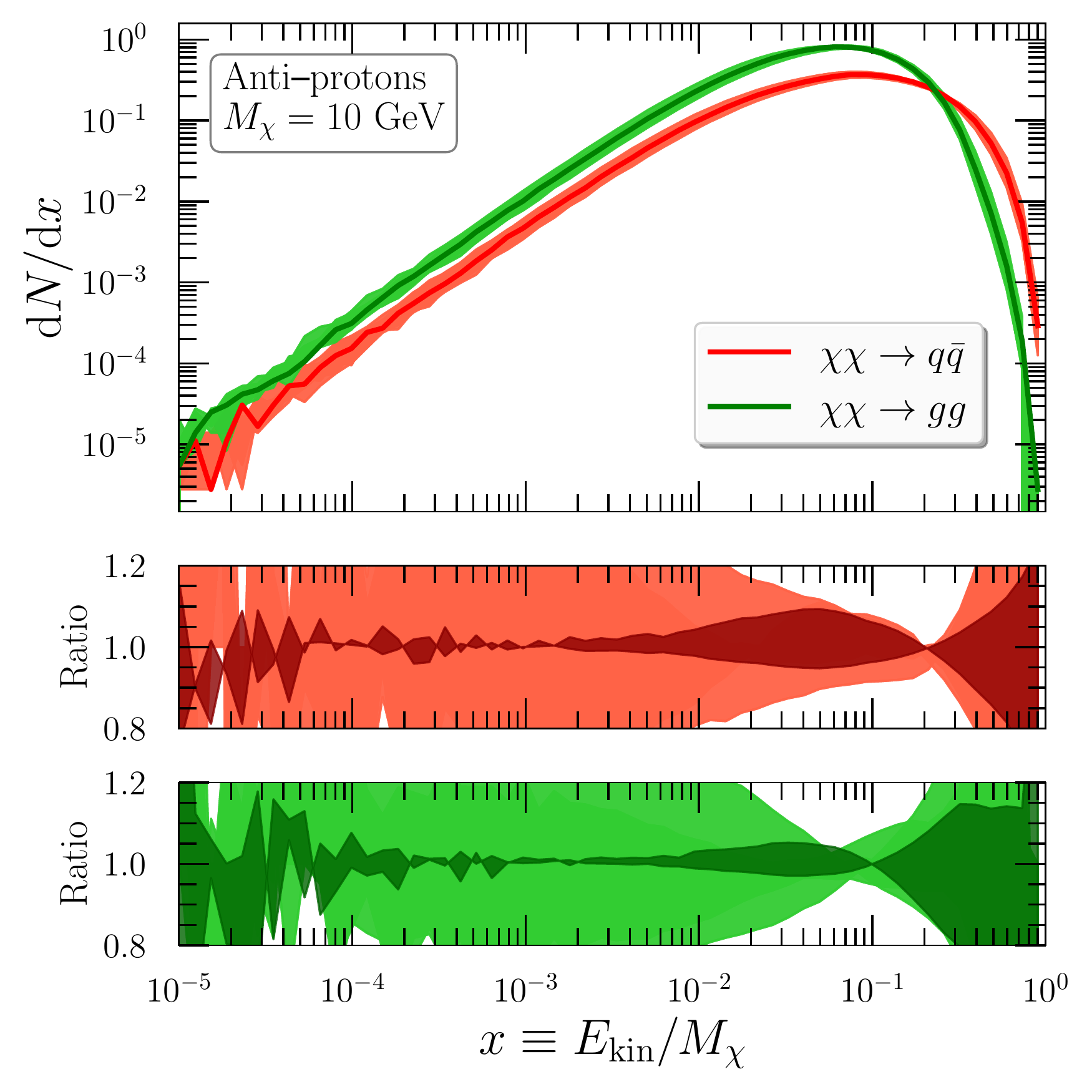}
    \hfill
    \includegraphics[width=0.49\linewidth]{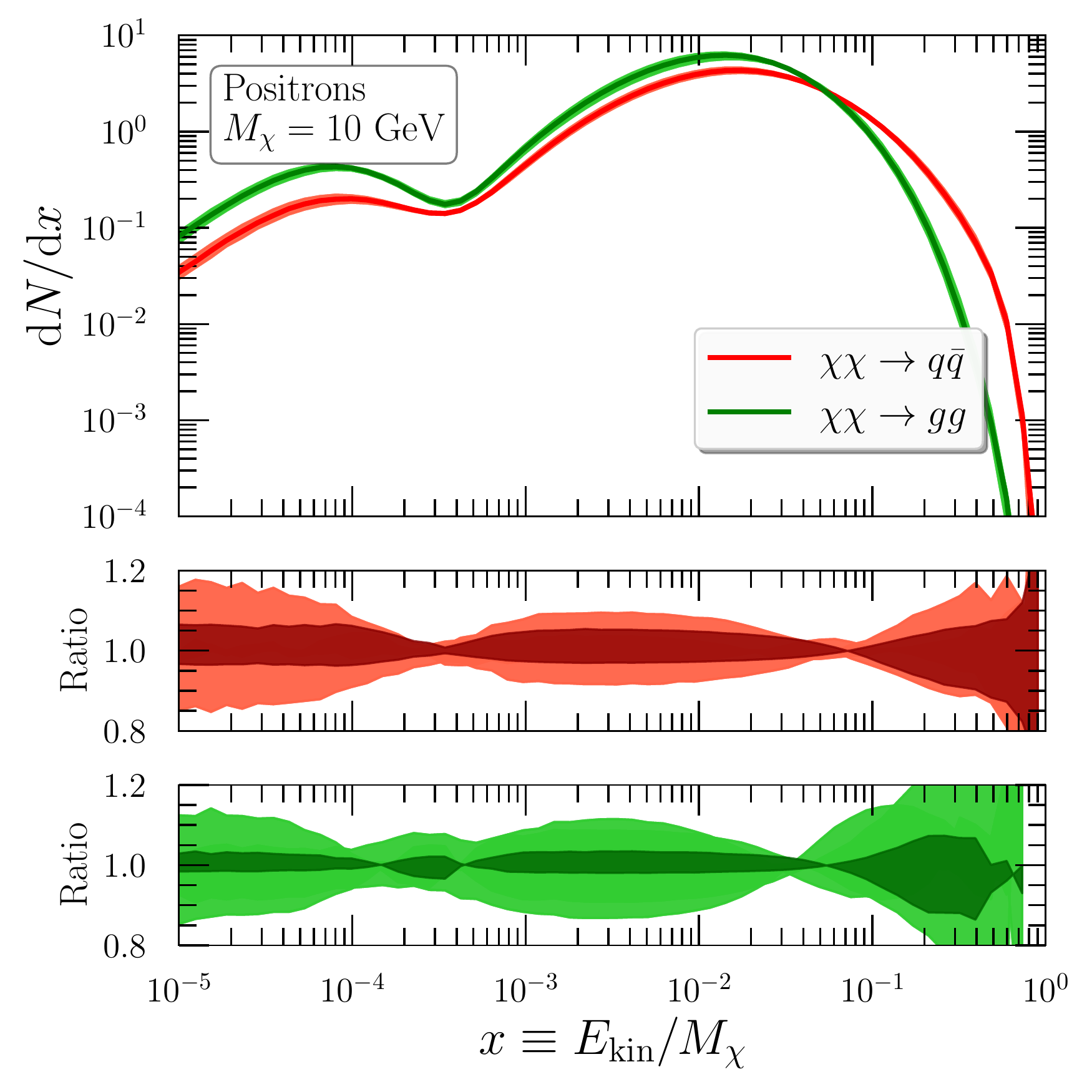}
    \vfill
    \includegraphics[width=0.49\linewidth]{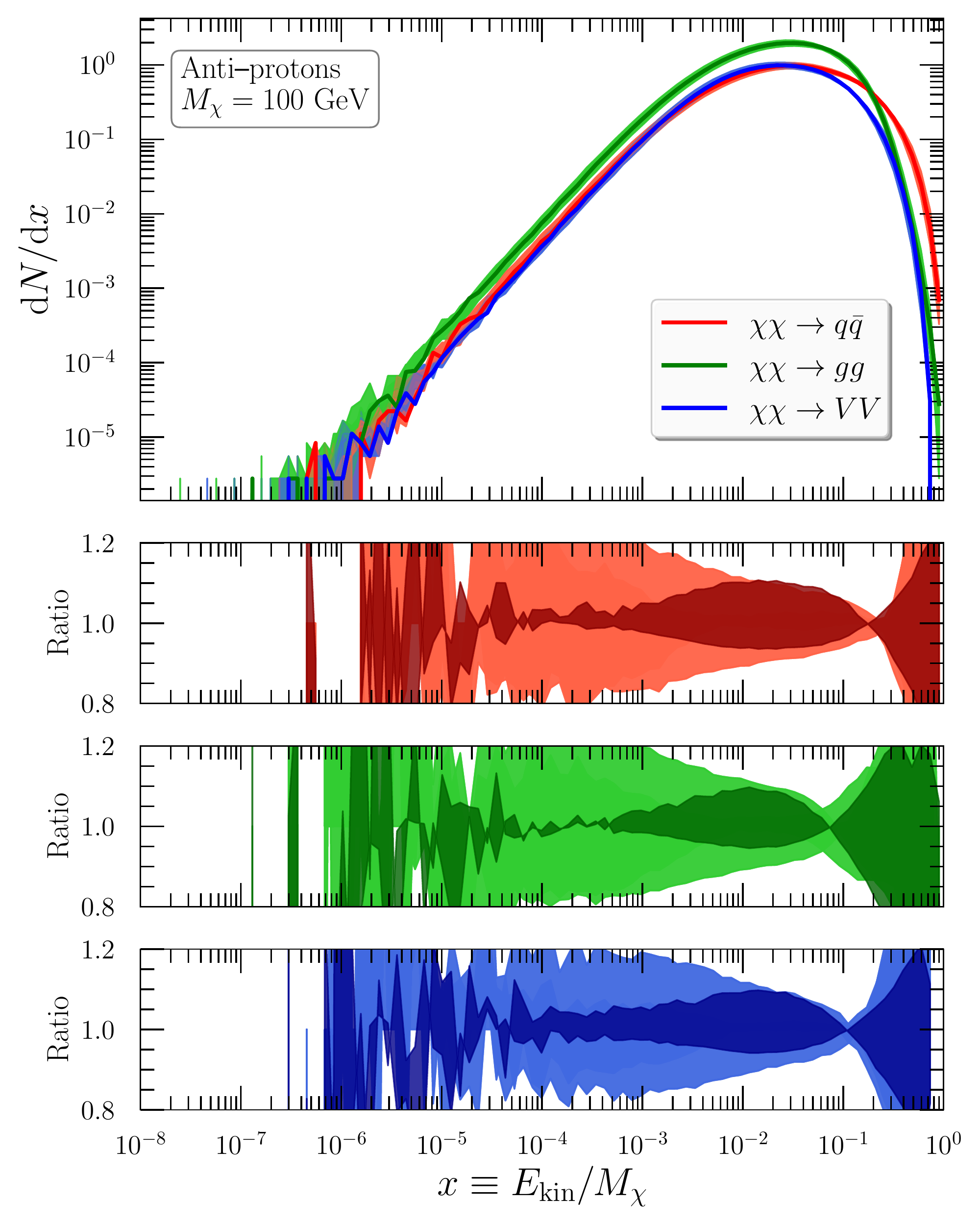}
    \hfill
    \includegraphics[width=0.49\linewidth]{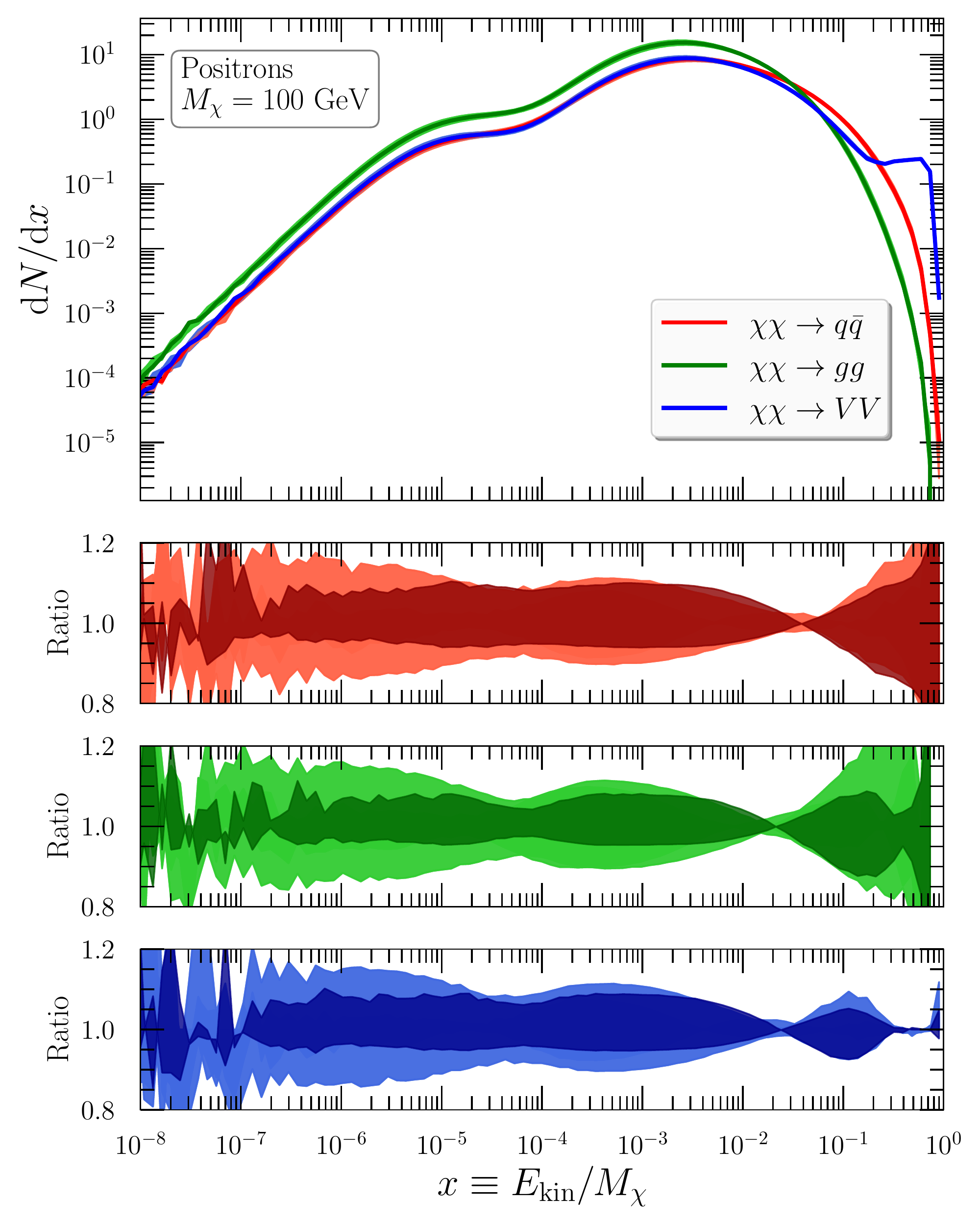}
    \caption{The scaled kinetic energy distribution of antiprotons ({\it left}) and positrons ({\it right}) in dark matter annihilation into $q\bar{q}$ (red), $gg$ (green) and $VV$ for dark matter mass of $10$ GeV (upper pane) and $100$ GeV (lower pane). For each pane, the dark shaded band corresponds to the parton-shower uncertainties while the light shaded band corresponds to hadronisation uncertainties.}
    \label{fig:spectra:mDM:10:100}
\end{figure}

\subsection{Assessing QCD uncertainties on antiproton spectra}

In this section, we quantify the impact of QCD uncertainties on particle spectra from dark matter annihilation for a few dark matter masses and annihilation channels. The results will be shown in the $x$ variable defined by 
\begin{eqnarray}
x \equiv \frac{E_{\rm kin}}{M_{\chi}} = \frac{E - m}{M_{\chi}},
\end{eqnarray}
with $E_{\rm kin}$ is the kinetic energy of the particle specie, $m$ is its mass and $M_{\chi}$ is the dark matter mass. We study the following annihilation channels: 
\begin{table}[!h]
    \centering
    \begin{tabular}{l l l}
    $M_\chi$          \hspace{2cm} &   $\chi \chi \to XX$ \hspace{2cm} & Spectra  \\
    $10~{\rm GeV}$     &   $q\bar{q}$, $gg$ & Figure \ref{fig:spectra:mDM:10:100} \\
    $100~{\rm GeV}$   &   $q\bar{q}$, $gg$, $VV$ & Figure \ref{fig:spectra:mDM:10:100} \\
    \end{tabular}
\end{table} \\
For the $q\bar{q}$ annihilation channel, we assume that the dark matter is annihilated to all the quarks except the top quark with ${\rm BR}(\chi \chi \to q\bar{q}) = 0.2, q=u,d,s,c,b$. For the $VV$ channel, we include both the $ZZ$ and $W^+ W^-$ channels with equal probabilities: {\it i.e.} ${\rm BR}(\chi \chi\to W^+ W^-) = {\rm BR}(\chi \chi \to ZZ) = 0.5$. The impact of QCD uncertainties on the antiproton spectra are shown in figure \ref{fig:spectra:mDM:10:100}. For comparison, we show the spectra of positrons as well. We can see that the QCD uncertainties resulting from parton-shower variations are subleading for dark matter mass of $10$ GeV and especially in the anti-matter spectra. As far as we go to high dark matter masses, for example $100$ GeV, these uncertainties become more competitive with the hadronisation uncertainties and reach up to $15\%$ in the peak region. The hadronisation uncertainties on the antiproton spectra are very important and can reach up to $20\%$ in the low energy region and about $10\%$ in the peak region. In the high energy region, both the perturbative and hadronisation uncertainties are important with the latter are dominant with respect to the former (and can reach up to $50\%$. Note that the position of the peak changes for some particle species. There are regions where all the variations result in no uncertainty at all, {\it e.g.} $x \simeq 0.2$ in the antiproton spectra in the $q\bar{q}$ final state. 

\section{Conclusions}
\label{sec:conclusions}
In this talk, we discussed the study of the QCD uncertainties on antiproton spectra from DM annihilation which we studied for the first time in \cite{Amoroso:2018qga}. We first discussed the physics modeling of antiproton production from dark matter annihilation and the minimal set of constraining data. We then performed several retunings of the Lund fragmentation function parameters in \textsc{Pythia}~8. We provided a minimal set of variations on the parameters that define a conservative estimate of the QCD uncertainties allowed by data. We have finally shown quantitatively the impact of the QCD uncertainties on the spectra of antiprotons from DM annihilation into $q\bar{q}$, $gg$, and $WW+ZZ$. Full data tables which can be used to update those in the PPPC4DMID are public can be found on \texttt{GitHub} and can be found in \url{https://github.com/ajueid/qcd-dm.github.io.git}. The tables can also be found in the next releases of \textsc{DarkSusy}~6 \cite{Bringmann:2018lay} and \textsc{MicrOmegas}~5 \cite{Belanger:2018ccd}.

\section*{Acknowledgements}

The work of AJ is supported in part by a KIAS Individual Grant No. QP084401 via the Quantum Universe Center at Korea Institute for Advanced Study. The work of JK is supported by the NWO Physics Vrij Programme ``The Hidden Universe of Weakly Interacting Particles" with project number 680.92.18.03 (NWO Vrije Programma),
which is (partly) financed by the Dutch Research Council (NWO). R. RdA acknowledges the Ministerio de Ciencia e Innovación (PID2020-113644GB-I00). PS is funded by the Australian Research Council via Discovery Project DP170100708 — ``Emergent Phenomena in Quantum Chromodynamic''. This work was also supported in part by the European Union’s Horizon 2020 research and innovation programme under the Marie Sklodowska-Curie grant agreement No 722105 — MCnetITN3.

\bibliographystyle{JHEP}
\bibliography{bibliography.bib}

\end{document}